# Rotational state-selection and alignment of chiral molecules by electrostatic hexapoles

Federico Palazzetti[a,*], Andrea Lombardi[a], Masaaki Nakamura[b], Shiun-Jr. Yang[b]
Toshio Kasai[b], King-Chuen Lin[b], Po-Yu Tsai[c], Dock-Chil Che[d]

[a]Università degli Studi di Perugia, Dipartimento di Chimica, Biologia e Biotecnologie, via Elce di Sotto 8, 06123 Perugia, Italy
[b]National Taiwan University, Department of Chemistry, Roosevelt Road, 10617, Taipei, Taiwan.
[c]Osaka University, Department of Chemistry, Toyonaka, Osaka, Japan.
[d]National Chung Hsing University, Taichung, Taiwan.
*Corresponding author: federico.palazzetti@unipg.it

**Abstract.** Electrostatic hexapoles are revealed as a powerful tool in the rotational state-selection and alignment of molecules to be utilized in beam experiments on collisional and photoinitiated processes. In the paper, we report results on the application of the hexapolar technique on the recently studied chiral molecules propylene oxide, 2-butanol and 2-bromobutane, to be investigated in selective photodissociation and enantiomeric discrimination.

## 1. INTRODUCTION

Electrostatic hexapoles [1-4] have been introduced several year ago and employed in the alignment and rotational state selection of molecules of increasing complexity (for alignment in gaseous streams see reference [5], in particular oxygen[6], benzene[7] and vortices[8]). They were initially applied to linear and symmetric-top molecules, while more recently they have been successfully applied to molecules of higher complexity such as the chiral molecules propylene oxide[9,10] (for a characterization of the reaction pathways of propylene oxide see reference [11]) and 2-bromobutane[12] (see also reference [13]). Hexapolar technique has been also employed for the first time in the selection of conformers of 2-butanol[14]. An important development is the combination of hexapolar electric fields (non-homogenous) with DC (Direct Current) orienting fields (homogenous) in order to obtain molecular orientation. This phenomenon consists in the polarization of the rotational angular momentum with respect a laboratory-fixed axis; it presents the great benefit of revealing geometrical features of molecules, which would be concealed in randomly rotating molecules. Computer simulations shown that oriented chiral molecules can give different results according to their mirror forms. Enantiospecific processes were predicted by molecular dynamics simulations of elastic collisions between oriented $H_2O_2$ and $H_2S_2$, some of the simplest chiral molecules[15] (for the characterization of the potential energy surfaces with rare-gas-atoms see references [16, 17]). Vector correlation also shown that the photodissociation by linearly polarized radiations of oriented chiral molecules can lead to different angular distributions of photofragments according to specific enantiomer. [18]

In this paper, we review the most important features of the hexapolar technique and show three case studies of application to three different chiral molecules: propylene oxide, 2-butanol and 2-bromobutane

## 2. BACKGROUND

### 2.1. Technical Aspects

Electrostatic hexapole is composed by six rods in aluminum, of length typical included between 50 cm to 2 m. In Figure 1a, we report a scheme of the experimental apparatus used to align propylene oxide[9] and 2-butanol[14].

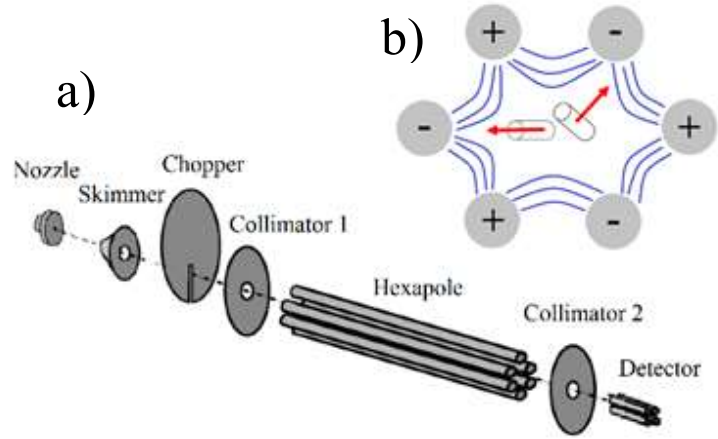

**FIGURE 1.** a) A typical scheme of the experimental setup. The molecular beam is introduced in the vacuum chamber through the nozzle; the skimmer makes a first selection of the direction, by excluding all the molecules, whose direction is not parallel to the molecular beam axis. The chopper is used for time-of-flight measurements and the collimator further select the molecular direction. After the hexapole, a second collimator selects the direction of the molecules, which eventually reach the detector, in this case represented by a quadrupole mass spectrometer. B) A view of the hexapole through the molecular-beam axis. The grey circles represent the rods, whose charges have opposite sign, while the blue curves indicate the strength line of the electric field. The cylinders and the red arrows represent two limit cases of distribution of the electric charge (the red arrow indicates the permanent dipole moment) with respect the molecular axis of a hypothetical prolate-top molecule (the cylinder). The picture in the left shows the permanent dipole moment parallel to the molecular axis, while in the picture in the right we illustrate of the perpendicular arrangement. (Adapted from references [9, 14]).

The six rods of the hexapole are parallel to the molecular beam axis and set at the same distance from this latter. A voltage, whose typical value is of some thousands of volt, is applied to the rods according to the scheme that adjacent rods possess opposite electrical charge (see Figure 1b) producing a non-uniform electric field that increases by approaching the rods. Molecules having a non-zero permanent dipole moment, experience an electric field $E$, whose intensity is given by the following equation:

$$E = 3V_0 \frac{r^2}{R^3} \qquad (1)$$

where $V_0$ is the voltage, $r$ is the distance of the molecule from the molecular beam axis and $R$ is the minimum distance of the surface of the rods from the molecular beam axis. Figure 1b shows a scheme of the molecular alignment of a hypothetical prolate-top molecule, where the permanent dipole moment is parallel or perpendicular to the molecular axis.

The molecules undergo a force, whose magnitude as a function of $r$, is

$$F(r) = -\frac{dW_{Stark}}{dr} = \frac{dW_{Stark}}{dE}\frac{dE}{dr},$$

where $dW_{Stark}$ is the Stark energy, *i. e.* the variation of the energy level in the electric field. A positive sign of the Stark energy, or equivalently a negative sign of the force, leads to a focusing trajectory in the hexapole field. On the contrary, a positive sign of the force indicates non-focusing trajectories.

## 2.2. Theoretical background and trajectory simulations

Asymmetric-top molecules are characterized by three different components of the moment of inertia along the principal axes, indicated by *a*, *b* and *c*. The rotational states are expressed by the quantum numbers $J$ and $M$, corresponding to the rotational angular momentum, $J$, and to its projection along a space-fixed quantization axis, $M$, i. e. the direction of the electric field or the molecular beam axis. The pseudo-quantum number $\tau$, replaces the quantum number $K$, which represents, for symmetric-top molecules, the projection of $J$ on the molecular axis; it is given by the combination of $K_{+1} + K_{-1}$, which represent the quantum number $K$ at the prolate-top and at the oblate-top limit, respectively. The molecular shape is commonly described by the Ray's parameter $k=(2B-A-C)/(A-C)$, where $A$, $B$ and $C$ are the rotational constants. The $k$ parameter is included between -1 and +1; it is equal to -1 in case of prolate-top molecules and +1 for oblate-top molecules. A complementary description of the molecular shape is given by the deformation indices, defined by means of the kinematic invariants. [19,20]

Trajectory simulations aim at reproducing the experimental focusing curve, that is a plot of the beam intensity as a function of the hexapole voltage, by calculating the possible paths of the molecule in the experimental apparatus. Input data are the molecular properties such as the components of the dipole moment and the Stark energy of the considered rotational levels, the geometrical features of the experimental apparatus, the hexapolar electric field and the beam conditions, such as the velocity distribution and the beam profile. Population distribution of the rotational states and conformers, when more than one is present, are calculated by making use of the Boltzmann distribution equation. Rotational and vibrational temperature, employed to calculate the population distribution of the rotational levels and conformers, respectively, are inferred by the best-fit between experimental and predicted focusing curves.

## 3. EXAMPLES AND DISCUSSION

*Propylene oxide.* Propylene oxide is an asymmetric-top molecule with a Ray's parameter of -0.87 and a permanent dipole moment of 2.2 D. Its focusing curve was characterized as a pure beam, seeded in He (20 % in propylene oxide) and seeded in Ar (20 % in propylene oxide). Details of the experimental apparatus and of trajectory simulations are given in references [9, 10]. In Figure 2 (left panel) we report, as an example, the experimental focusing curve of pure propylene oxide and the simulated ones by assuming adiabatic and non-adiabatic transitions in case of avoided curve crossing.

*2-butanol.* This molecule presents a complex conformer manifold because of the torsion around the second and the third carbon, which originates rotamers T, G+ and G-, and the torsion around the second carbon and the oxygen, originating rotamers t, g+ and g-; the resulting conformers (nine) are given by the combination of these rotamers (for details see reference [14]). Each conformer possesses characteristic rotational constants, dipole moments and related components, determining different aligning process, as evidenced by the focusing curves (Figure 2, central panel) of the nine conformers of the molecular beam seeded in He (5 % in 2-butanol).

*2-bromobutane.* Differently from the previous two chiral molecules, 2-bromobutane have been also oriented (for details of the experimental setup see reference [12]). The traditional ion lens setup, that makes use of the extraction field for the molecular orientation, has been modified, by introducing an independent orienting field, consisting of a pulsed voltage of very short response time, 30 ns, applied at the ion extraction stage, that permits to change intensity and sense of the orienting field. Figure 2 (right panel) shows the different intensity of the bromine fragment with and without the application of the pulsed voltage as a function of the time-of-flight, on a prealigned molecular beam by a hexapolar field of 4.0 kV/cm. In this conditions, the Br atom in 2-bromobutane, can be displaced along the time-of-flight axis, toward the detector (forward-orientation) or toward the nozzle (backward orientation). The fastest peak is ascribed to $^{79}$Br dissociated from the forward oriented 2-bromobutane, while the slower one to $^{81}$Br dissociated from the backward oriented 2-bromobutane. The central peak is due to the overlap of $^{79}$Br from the backward oriented molecule and $^{81}$Br from the forward oriented one. The pulsed voltage is directed along the time-of-flight axis and determines a backward orientation of the molecules, which results in the increasing of the signal attributed to the backward oriented molecules. 2-bromobutane was the first chiral molecule to be oriented and the related photodissociation dynamics study is reviewed in an accompanying paper.[21]

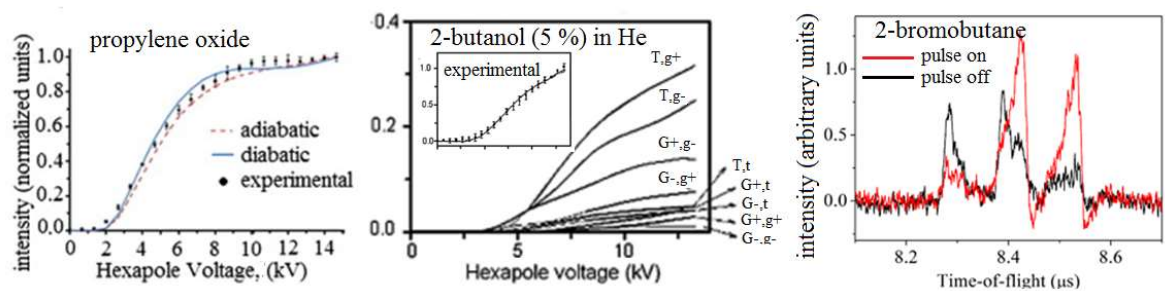

**FIGURE 2.** The focusing curves of pure propylene oxide molecular beam (left panel) and of the conformers of 2-butanol seeded in He (central panel), in the inset we show the total focusing curve obtained by experiment (dots) and simulation (line). Figure 2c shows the change of intensity of detected Br photofragment with (red line) and without (black line) application of the orienting field. (Adapted from references [9, 12, 14])

## ACKNOWLEDGMENTS

FP, AL and VA acknowledge the Italian Ministry for Education, University and Research (MIUR) for financial support ("Scientific Independence for young Researchers", SIR 2014, RBSI14U3VF).